# Super-FRS GEM-TPC Prototype Development Based on n-Xyter Asic for the FAIR Facility

F. García*[1], R. Turpeinen[1], R. Lauhakangas[1], E. Tuominen[1], R. Janik†[2], P. Strmen[2], M. Pikna[2], B. Sitar[2], B. Voss[3], J. Kunkel[3], V. Kleipa[3], A. Prochazka[3], J. Hoffmann[4], I. Rusanov[4], N. Kurz[4], S. Minami[4]

[1]Helsinki Institute of Physics, University of Helsinki, 00014 Helsinki, Finland
[2]FMFI Bratislava, Comenius University, Bratislava, Slovakia
[3]Detector Laboratory, GSI Helmholtzzentrum, Darmstadt 64291, Germany
[4]Experimental Electronics Department, GSI Helmholtzzentrum, Darmstadt 64291, Germany

*Abstract*–**The FAIR[1] facility is an international accelerator centre for research with ion and antiproton beams. It is being built at Darmstadt, Germany as an extension to the current GSI research institute. One major part of the facility will be the Super-FRS[2] separator, which will be include in phase one of the project construction. The NUSTAR experiments will benefit from the Super-FRS, which will deliver an unprecedented range of radioactive ion beams (RIB). These experiments will use beams of different energies and characteristics in three different branches; the high-energy which utilizes the RIB at relativistic energies 300-1500 MeV/u as created in the production process, the low-energy branch aims to use beams in the range of 0-150 MeV/u whereas the ring branch will cool and store beams in the NESR ring. The main tasks for the Super-FRS beam diagnostics chambers will be for the set up and adjustment of the separator as well as to provide tracking and event-by-event particle identification. The Helsinki Institute of Physics, the Comenius University, and the Detector Laboratory and Experimental Electronics at GSI are in a joint R&D phase of a GEM-TPC detector which could satisfy the requirements of such diagnostics and tracking chambers in terms of tracking efficiency, space resolution, count rate capability and momenta resolution. The current status of the first prototype and the preliminary results from the test beam campaign S417 using the n-Xyter chips mounted on GEMEX cards will be shown.**

## I. Introduction

THE FAIR facility at GSI will stand for the research with ion and antiproton beams, in cooperation with their users and the international community. With the new project, GSI aims to provide scientists in Europe and the world with an outstanding accelerator and experimental facility for studying matter at the level of atoms, atomic nuclei, protons and neutrons as the building blocks of nuclei - and part of a wider family called hadrons - and the subnuclear constituents called quarks and gluons.

The facility will provide an extensive range of beams from protons and antiprotons to ions up to uranium with world record intensities and excellent beam quality in the longitudinal as well as transverse phase space. The scientific goals pursued at FAIR[3] include:

- Studies with beams of short-lived radioactive nuclei, aimed at revealing the properties of exotic nuclei, understanding the nuclear properties that determine what happens in explosive processes in stars and how the elements are created, and testing fundamental symmetries.
- The study of hadronic matter at the subnuclear level with beams of anti-protons, in particular of the following key aspects: the confinement of quarks in hadrons, the generation of hadron masses by spontaneous breaking of chiral symmetry, the origin of the spins of nucleons, and the search for exotic hadrons such as charmed hybrid mesons and glueballs.
- The study of compressed, dense hadronic matter through nucleus-nucleus collisions at high energies.
- The study of bulk matter in the high density plasma state, a state of matter of interest for inertial confinement fusion and for various astrophysical sites.
- Studies of Quantum electrodynamics (QED), of extremely strong electromagnetic.

The NuSTAR experiments will be dedicated to the study of Nuclear Structure, Astrophysics and Reactions. In particular with the use of beams of radioactive species separated and identified by the Superconducting Fragment Recoil Separator (Super-FRS). No experimental programme in nuclear physics stands entirely on its own and NuSTAR is not an exception. Strictly speaking some activities at GSI will fall naturally into the sphere of interest of NuSTAR, such as the search for new superheavy elements and the study of their physical and chemical properties. Since the success of the experimental programme of NuSTAR depends critically on the

Manuscript received November 16, 2012. This work was supported by the Ministry of Education of Finland.

F. García* is with the Helsinki Institute of Physics, University of Helsinki, P.O. Box 64, FI-00014 University of Helsinki, Finland (telephone: +358-9-19151086, e-mail: Francisco.Garcia@helsinki.fi).

R. Turpainen, E. Tuominen, R. Lauhakangas , are with the Helsinki Institute of Physics, University of Helsinki, P.O. Box 64, FI-00014 University of Helsinki, Finland (e-mails: Raimo.Turpeinen@helsinki.fi, Eija.Tuominen@helsinki.fi, Rauno.Lauhakangas@helsinki.fi).

R. Rudolf Janik, M. Pikna, B. Sitar, P. Strmen are with the FMFI Bratislava, Comenius University, Bratislava, 84248 Slovakia (e-mails: janik@fmph.uniba.sk, pikna@fmph.uniba.sk, sitar@fmph.uniba.sk, strmen@fmph.uniba.sk).

B. Voss, J. Kunkel, V. Kleipa, A. Prochazka are with the Detector Laboratory at GSI, 64291 Darmstadt, Germany (e-mails: B.Voss@gsi.de, j.kunkel@gsi.de, v.kleipa@gsi.de, a.prochazka@gsi.de).

J. Hoffmann, I. Rusanov, N. Kurz, S. Minami are with the Experimental Electronics Department at GSI, 64291 Darmstadt, Germany (e-mails: j.hoffmann@gsi.de, I.Rusanov@gsi.de, N.Kurz@gsi.de).

specifications and resulting properties of the Super-FRS, this device will be also seen as an integral part of its activities.

The fragment separator FRS provides high energy, spatially separated monoisotopic beams of exotic nuclei of all elements up to uranium[4,5]. The fragments are separated in flight, thus the accessible lifetimes are determined by the time-of-flight through the ion-optical system, which range from the submicrosecond domain upwards.

The FRS[4] has proven to be an extremely versatile instrument for new nuclear and atomic studies, as well as for experiments in applied physics. More than 150 new isotopes have been discovered and studied directly at the FRS, including the doubly magic nuclei $^{100}$Sn and $^{78}$Ni. Neutron-deficient projectile fragments between lead and uranium have been excited by the Coulomb field of heavy targets to investigate low-energy fission properties for a large number of nuclei for the first time.

Additionally, the FRS delivers fragment beams for decay and reaction studies to the ALADIN-LAND setup and the KAOS spectrometer. Fragments are also delivered to the ESR, for experiments such as precision mass and lifetime measurements. Moreover, the FRS has been used as a high-resolution magnetic spectrometer. Precise momentum measurements led to the discovery of new halo properties, the first observation of deeply bound pionic states in heavy nuclei, and several fundamental properties in atomic collision physics. In several fields of applied physics important studies were performed, such as the full characterization of heavy-ion therapy beams and optimization of PET diagnostics for the GSI cancer therapy project.

Although the FRS facility has contributed to the field of heavy-ion science with great success, there are four major enhancements that can be applied to improve the method considerably: 1. Higher intensity of the primary beams, 2. increased transmission for fission fragments produced by uranium projectiles, 3. improved transmission of fragments to the dedicated experimental areas, 4. larger acceptance of fragments by the storage cooler ring. These improvements are addressed in the recently proposed international rare-isotope facility at GSI, which includes a powerful driver accelerator, a large-acceptance superconducting in-flight separator, and a storage-cooler ring complex optimized to accept large-emittance secondary beams.

The Superconducting fragment Separator (Super-FRS)[5] is a powerful in-flight facility which will provide spatially separated isotopic beams up to elements of the heaviest projectiles. It is superior to the present FRS due to the incorporation of more separation stages and larger magnet apertures through the use of superconducting coils. The Super-FSR is based on results, experience, methods and techniques which were pioneered and developed for relativistic heavy ions at the FRS.

## II. Super FRS Diagnostic chambers

It is planned to implement a detection system that can be commonly used for all experiments at the different Super-FRS branches and comes with its associated data acquisition scheme. The main task of this combined system is threefold:

- it can be used to set up and adjust the separator,
- it provides the necessary measures for machine safety and monitoring,
- it allows for an event-by-event particle identification, tracking and characterization of the produced rare ion species.

Furthermore, the beam intensities at different locations in the separator are to be monitored, e.g. as to normalize measured rates to be able to extract absolute cross sections. The modi operandi depend strongly on these given tasks and the necessary requirements for the combined detector and acquisition systems will be given.

Setting up and adjusting the separator can be done at a low rate for almost any detector system. The main design goal is to get an easy to maintain, reliable system. An online monitoring has to be performed, especially in the target and beam catcher areas. Any deviation of the primary beam from its nominal position should lead immediately to an interlock condition. The main challenge is to cope with the very high intensities and background radiation here. The design of the detector systems should allow extended periods of operation without maintenance hands-on.

For almost all experiments, the separator is to be treated as first part of the experimental setup. The beam particles entering the different branches have to be identified and their longitudinal and transverse momentum components should be known. For tracking experiments to be carried out in the Low-Energy Branch and the High-Energy Branch, the measurement has to be performed on an event-by-event basis. This implies that the data acquisition system of the particular experiments and the Super-FRS should be closely coupled if not identical. The Super-FRS data taking will therefore be designed in accordance with the common NUSTAR data acquisition scheme. The requirements on detector systems are demanding at the entrance of the main separator, where rates up to $10^9$ particles /s can be expected.

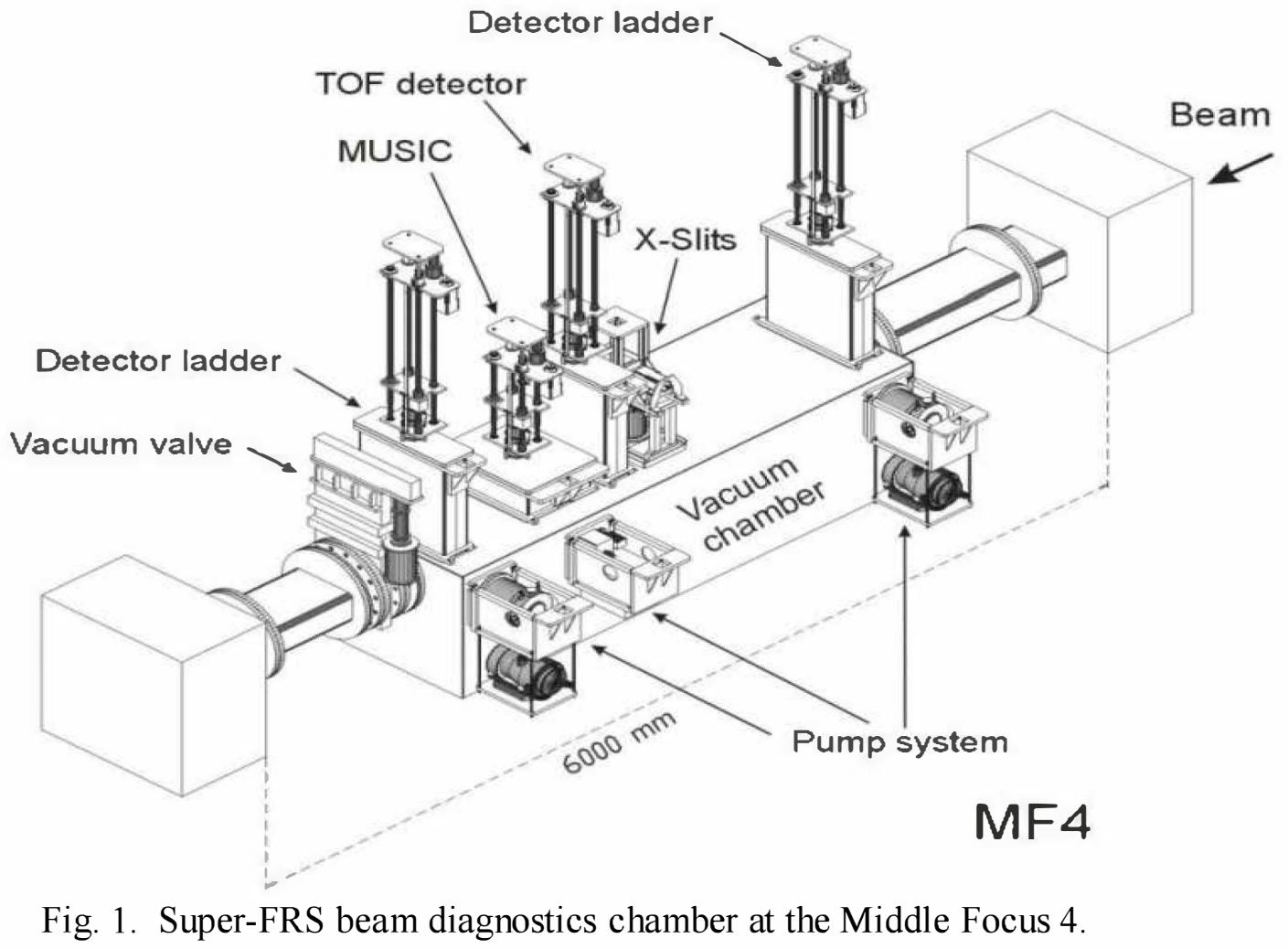

Fig. 1. Super-FRS beam diagnostics chamber at the Middle Focus 4.

The locations for the different detection systems are shown in Figure 1. It is foreseen to be able to run for about one year without opening the sections along the S-FRS. Generally we foresee UHV material although a pressure of $10^{-7}$ mbar is sufficient. The choice of the particular detector systems is driven by the idea trying to benefit from the various developments that are currently done in the detector laboratory and accelerator division.

## III. GEM-TPC DETECTOR DEVELOPMENT

The Diagnostics for slowly extracted beams are defined as slow extraction when the extraction times are above 100 ms. For this purpose Time Projection Chambers with Gas Electron Multipliers (GEM-TPC) will be installed and the requirements for such detectors are:

- No interference with the beam
- Large dynamic range
- Intensities less than 100 kHz

In order to satisfy these requirements a working group was formed, which produce several designs[6]. Afterwards was decided that a GEM stack will be added to the conventional TPC and in that way a first prototype was developed. This first GEM-TPC was read by delayed lines, and at the FRS the detector was tested in the beam campaign S363[7].

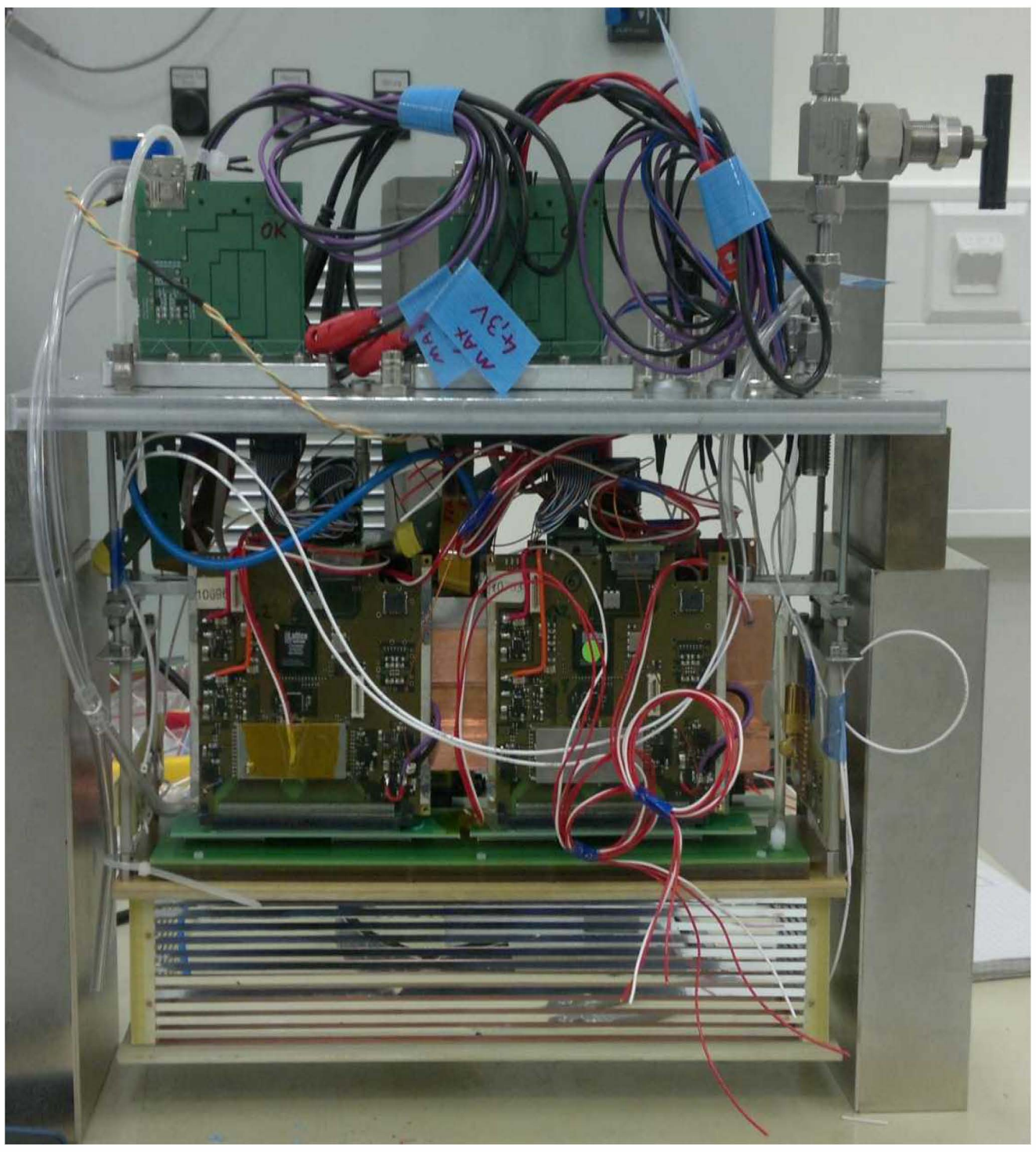

Fig. 2. Super-FRS GEM-TPC prototype HB3, equipped with four GEMEX cards.

During this year the beam campaign S417 gave us the opportunity to test a further evolution of the GEM-TPC, now equipped with GEMEX[8] readout cards, which each of them has two n-Xyter[9] chips. This prototype is called HB3, which means Helsinki-Bratislava prototype number. One GEMEX card can read up to 256 channels. Two GEMEX cards were used to read 512 strips located in one side of the pad plane and the other side with 512 were read by two other cards.

The readout board has a geometry defined by 512 strips cut in half, which are 200 μm in width and with a pitch of 500 μm. This type of readout geometry suits very well to cope with high count rate.

In the Fig. 3 is shown a close up of the readout strips and the footprint of the connector can be seen as well. Each footprint can have a 130 pin and Panasonic connectors were used, from which only 128 pins will be connected to readout channels and the rest two will be grounded.

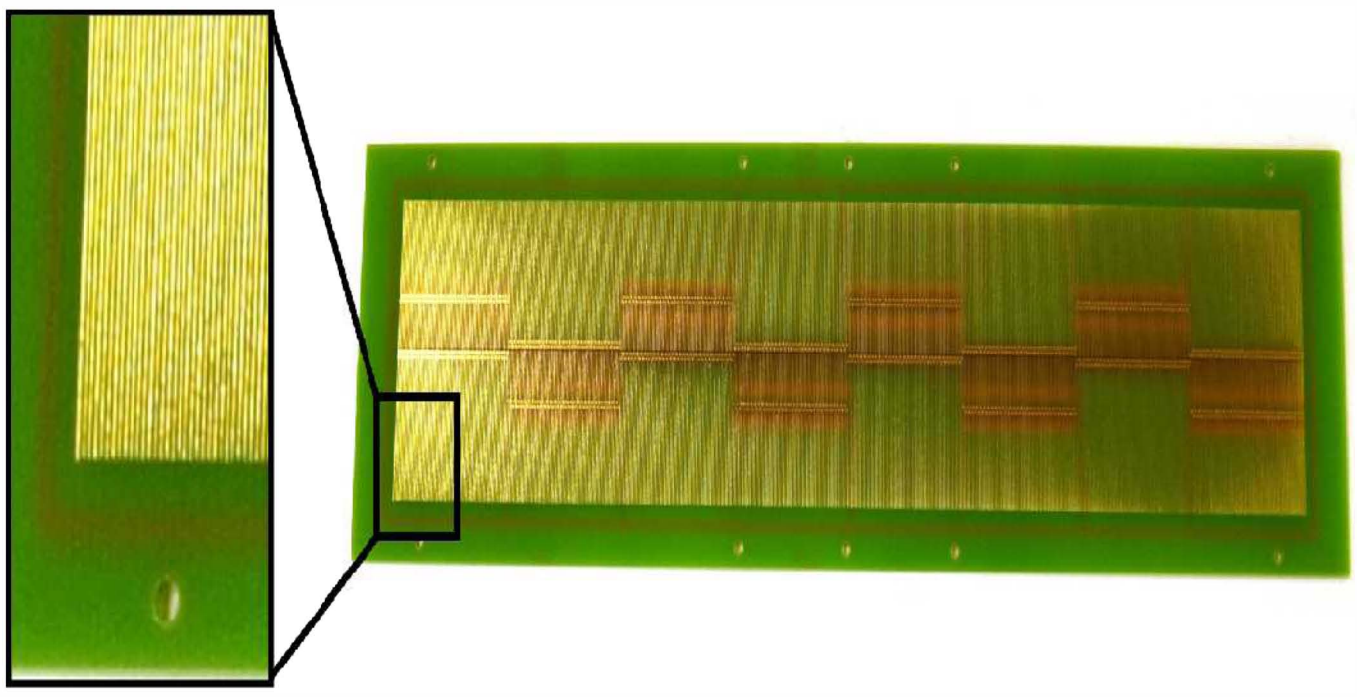

Fig. 3. Readout Pad plane. It consists of 512 strips cut in the middle, to create a total of 1024 channels.

Since the strips were cut in half, we obtained two back to back detectors parallel to the beam. Therefore the data was presented in HB3 front and HB3 back and each of them will have 512 channels.

## IV. RESULTS FROM TEST BEAM AT FRS

During the S417 campaign, we have got the opportunity to test our detector. One of the task was to test how the GEMEX cards with the n-Xyters will perform integrated to a GEM detector.

The primary beam particles were $^{197}$Au with an energy of 750 MeV/u and the intensity was varying around $10^7$ ions/spill. The spills has a duration of about 8 - 10 s and the beam was moved from the center to left to the right, in addition to that the beam was focused and defocused. Below will be shown some of the correlations plots between the HB3 and the TPC tracker. This allow us to get first results on the performance of the detector, nevertheless a detailed data analysis need to be carried out in order to extract the tracking efficiency and position resolution.

In the Fig. 4 the whole test beam geometry is shown, as it can be seen the HB3 was in between two conventional TPC

detectors, which were acting as a tracker. This is needed in order to later study in detail the HB3 performance.

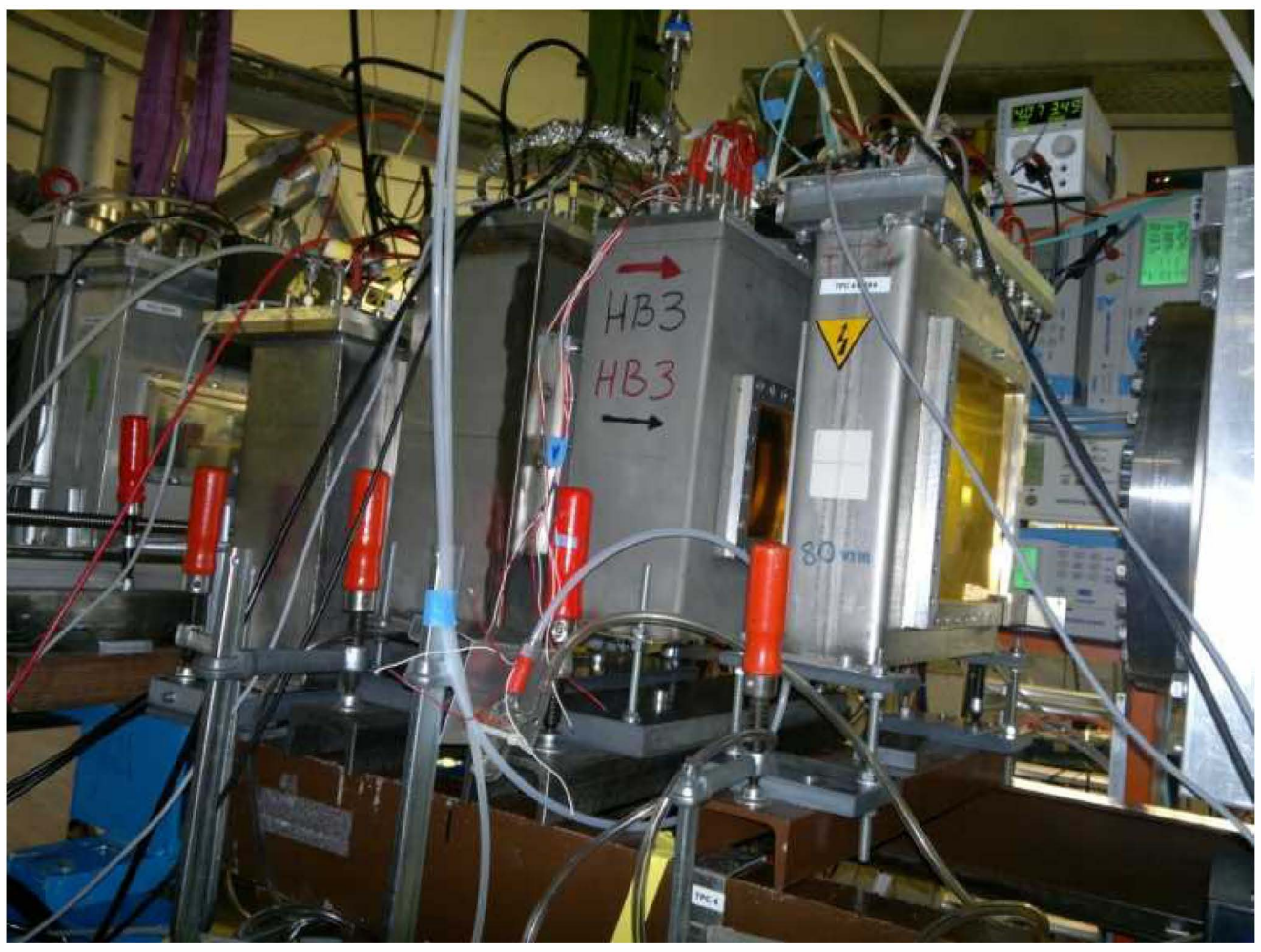

Fig. 4. SuperFRS GEM-TPC prototype HB3 between two conventional TPCs in the S2 cave at the FRS - GSI.

In order to assets the performance of the GEM-TPC, we will first show plots of the correlation between the tracker and the HB3. It is important to highlight that the HB3 will be treated as two detectors since we have the strips cut in half, therefore we will call to the half that will be hit by the beam in downstream direction first as the HB3 front and the subsequent one as HB3 back.

In the same manner; the results will be presented for the X-coordinate first and then for the Y-coordinate. It is important to stress that the X-coordinate is given by the physical location of the strips in the readout pad plane and the Y-coordinate by the drift time taken by the electrons moving from the location where the primary ionization occurred in the field cage towards the readout plane.

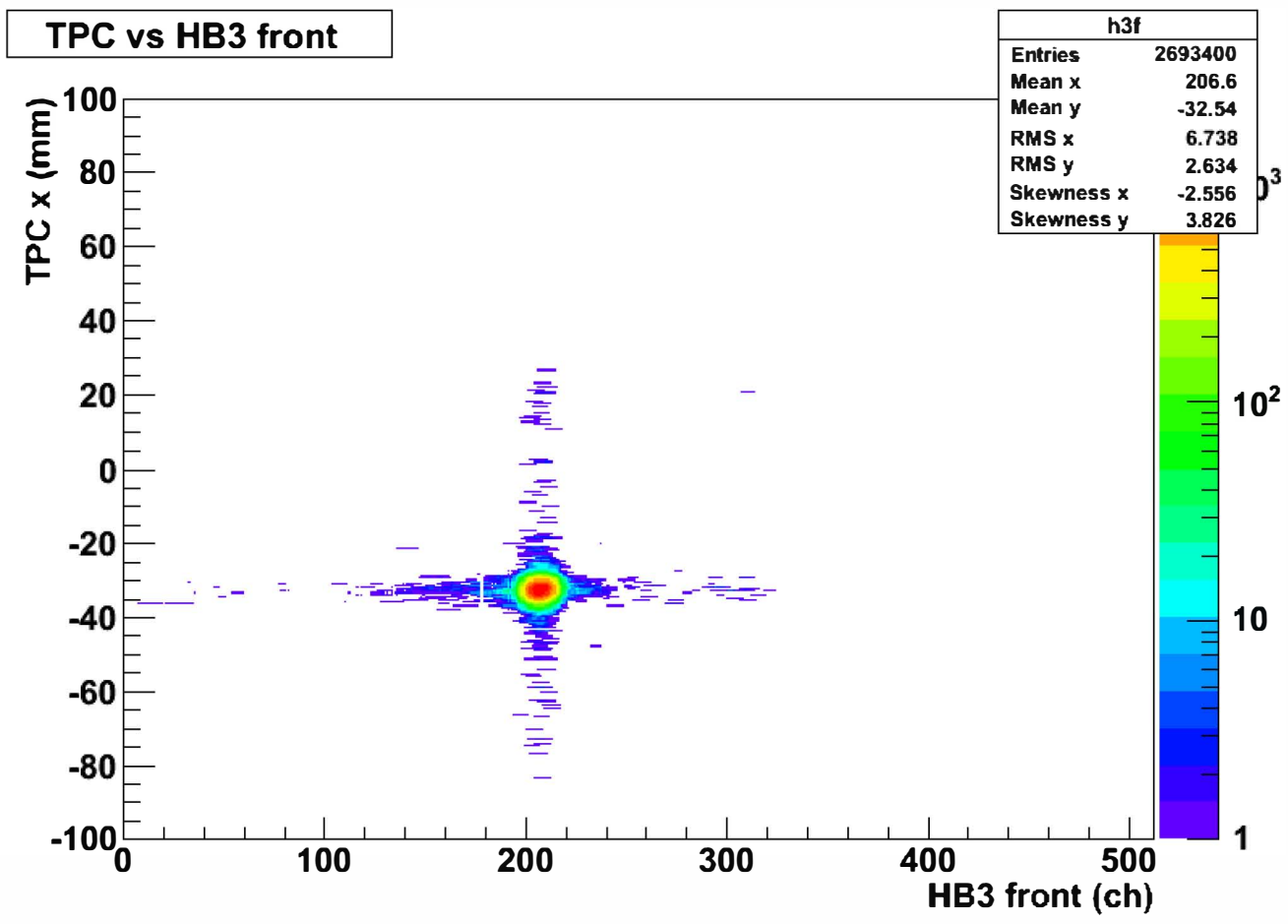

Fig. 5. Position correlation in X-coordinate between the HB3 front and TPCs tracker.

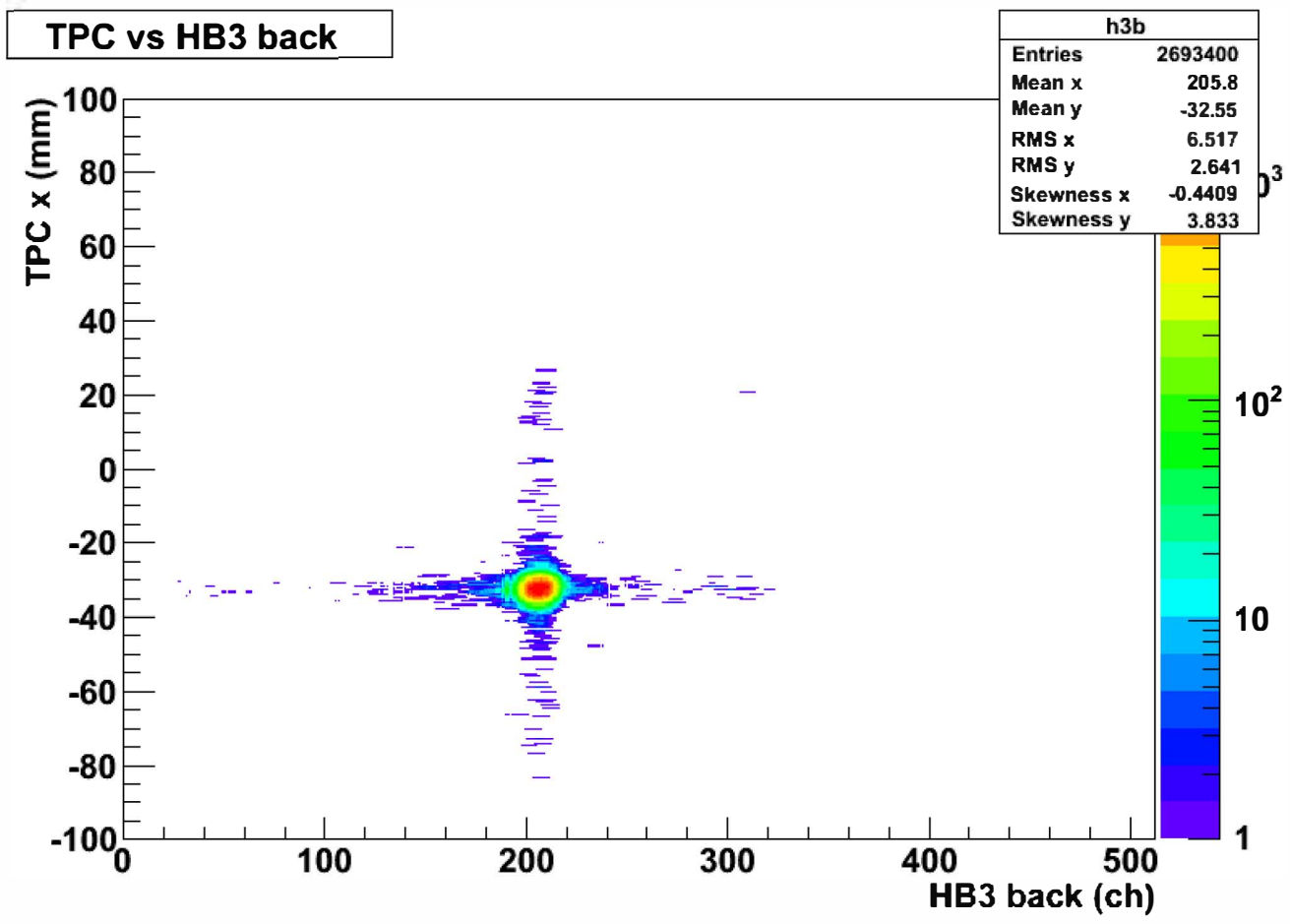

Fig. 6. Position correlation in X-coordinate between the HB3 back and TPCs tracker.

Above can be seen that the HB3 front and back perform very well and uniformly with respect to the TPC tracker (See Fig. 5 and 6). This will means that the distribution of hits registered by the tracker has a corresponded hit in the HB3 front and back.

Analogue distribution has been obtained in the Y-coordinated, which will be dominated by the drifted time along the field cage (see Fig. 7).

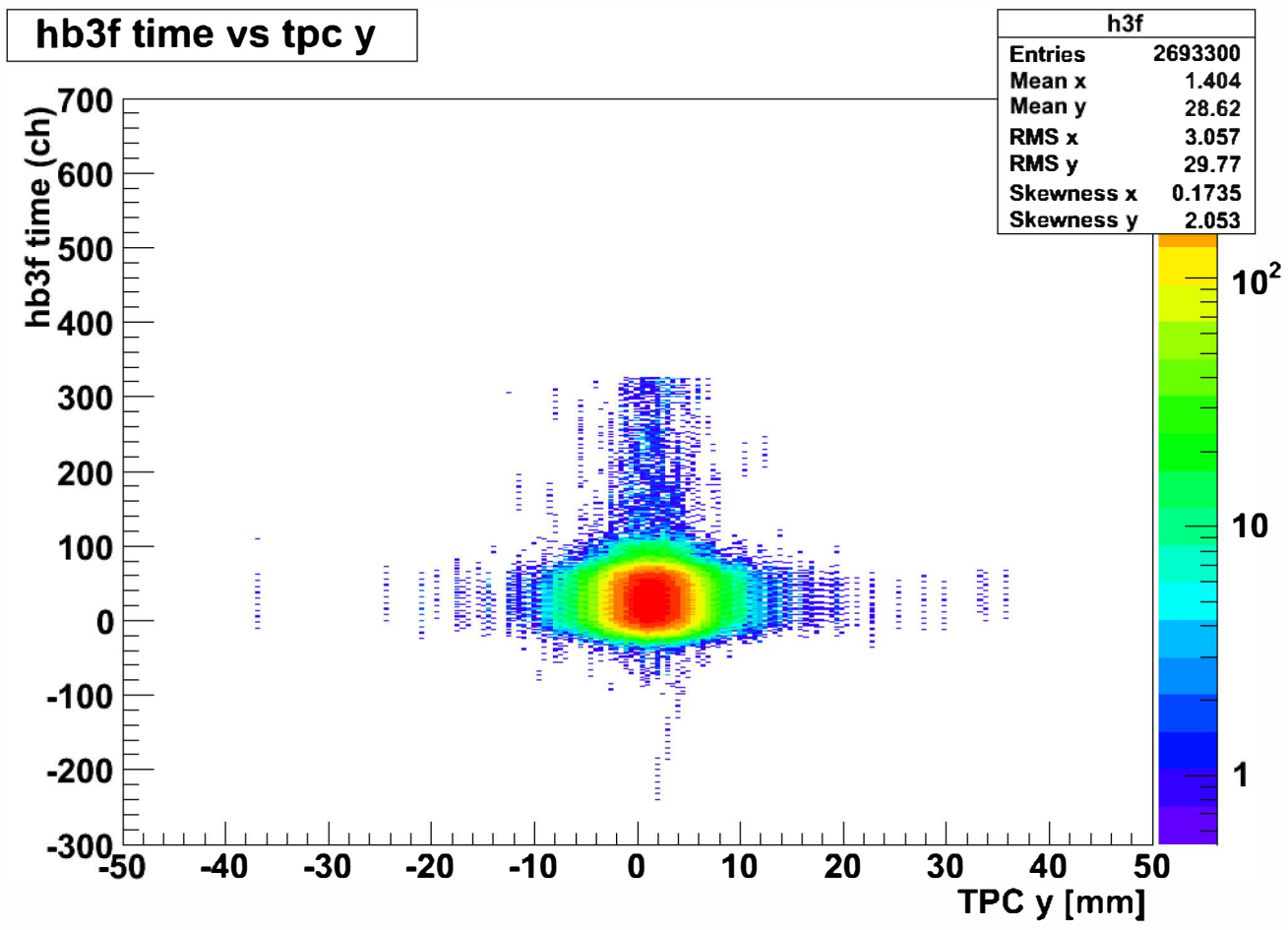

Fig. 7. Position correlation in Y-coordinate between the HB3 front and TPCs tracker.

We can observe that the Time distribution shown above match very well with the one provide by the tracker and in addition to that the correlation between the internal time counter and the time produced by the HB3 front is shown as well in Fig. 8.

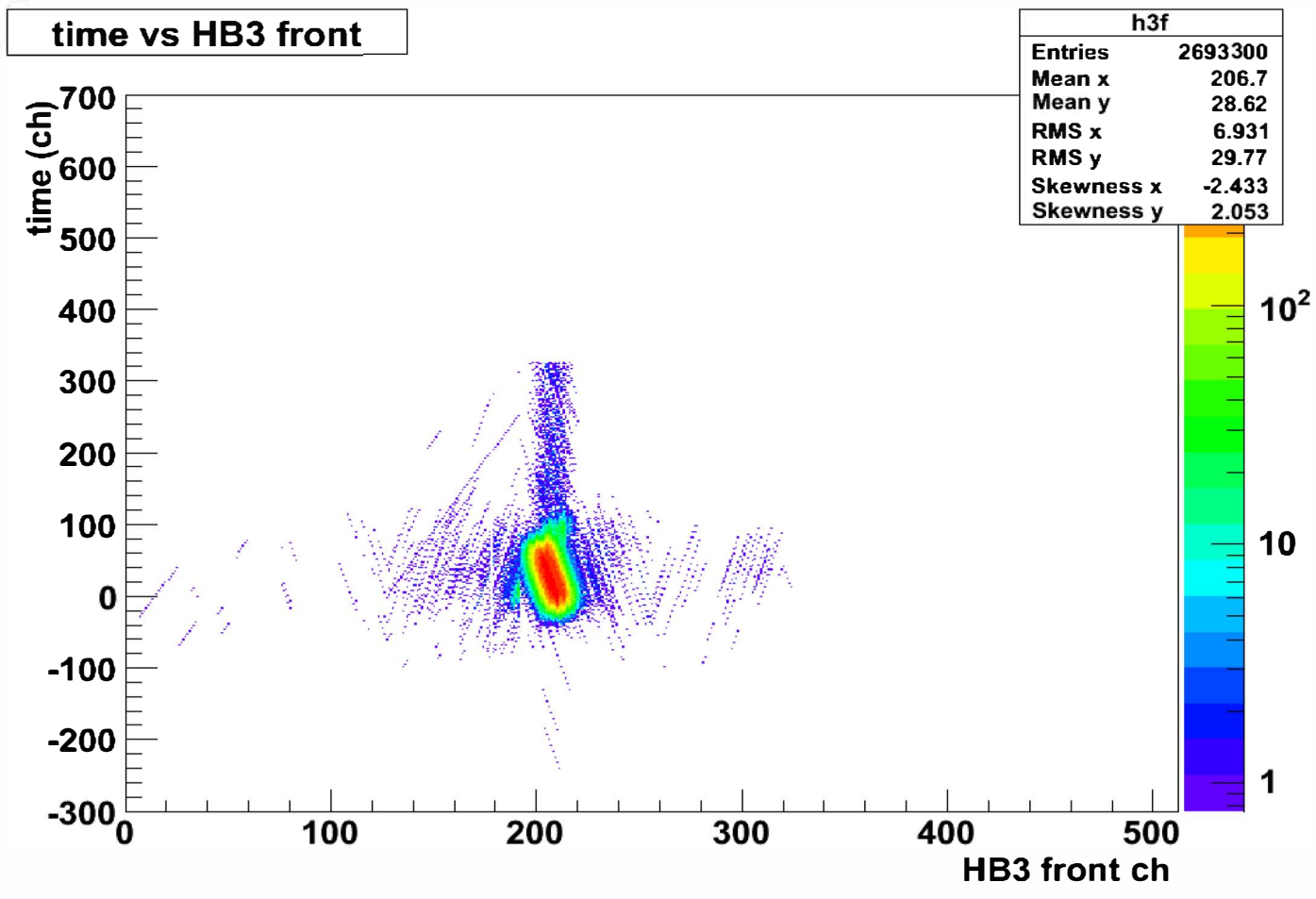

Fig. 8. Time correlation between the HB3 front and internal time counter.

Once we have found that the detectors are performed in a quite similar way, and the correlation between different parameters match quite well. One can finally take a look to the correlations between the front and back of the HB3 itself, and this is shown in the Fig. 9.

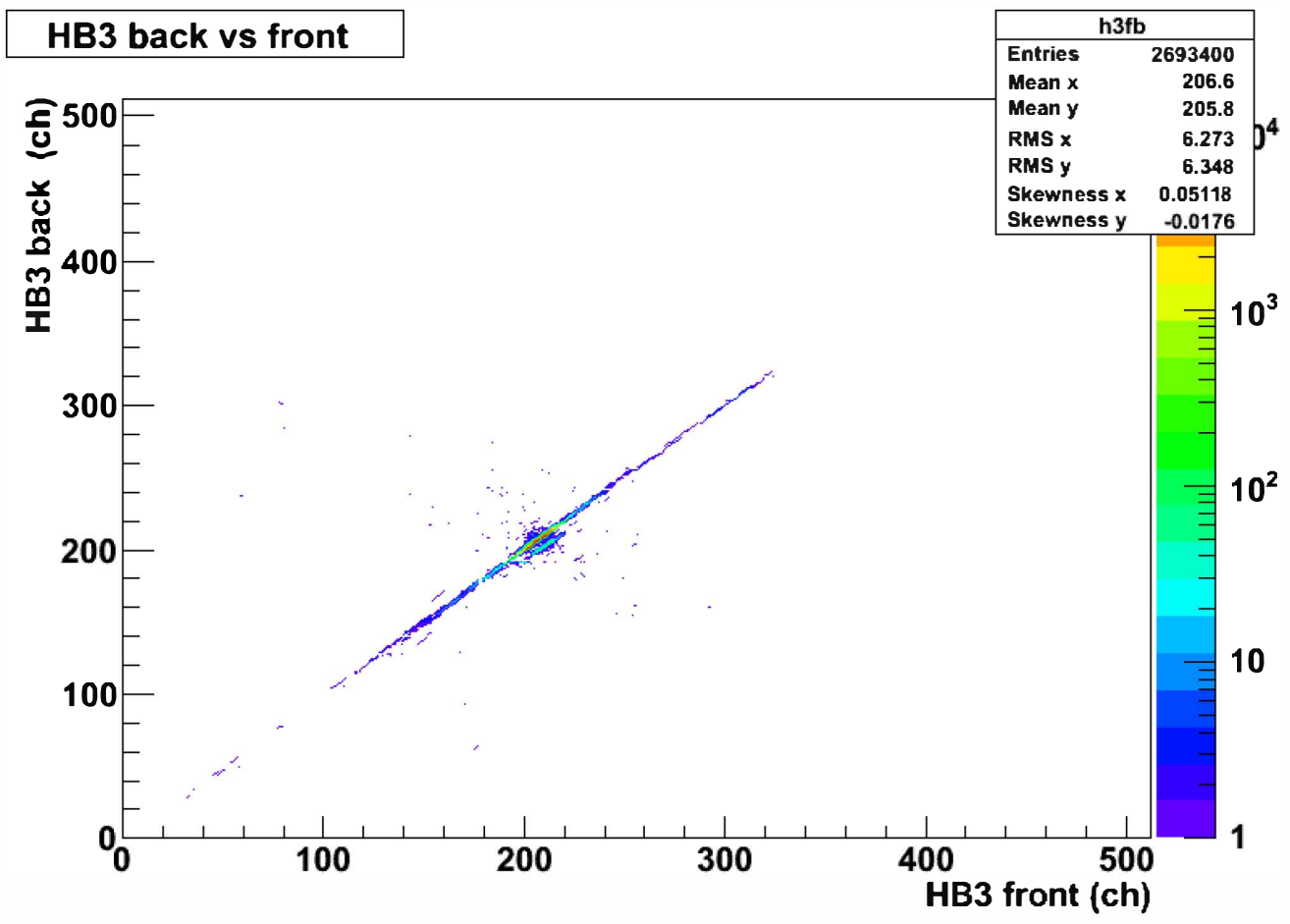

Fig. 9. Correlations of hits between the front and back of the HB3.

From the above presented plot one can conclude that both sides of the detector were performing quite similar and very linearly. This analysis by itself do not represent a final statement of the tests carry out at S417 campaign, but it is the first step towards a full analysis, now an effort has to be done, in order to find the position resolution and the tracking efficiency. These parameters will bring more detail into the performance of the HB3, since we will need to constrain the hits in the HB3 with the extrapolated track position from the conventional TPC. Moreover we will find finally what was the strip multiplicity for the clusters, since a cauterization algorithm has to be created.

## V. CONCLUSIONS

Results from the test Beam campaign S417 has shown that the third prototype of a GEM-TPC for the SuperFRS has been tested and that the concept of using a GEM has been proven. The next step is to carry out a further analysis of the data in order to find the position resolution and tracking efficiency and this will demand a bigger effort.

## ACKNOWLEDGMENT

We will want to acknowledgment to the Finnish Ministry of Education for the long term funding of the Fair facility participation.